\documentclass[aps,twocolumn,floats,PhysRevD,nofootinbib,groupedaddress,10pt]{revtex4-1}
\usepackage[utf8]{inputenc}

\usepackage{graphicx,amsmath,amsfonts,amssymb,slashed}

\usepackage{amsmath,amsfonts,euscript,hyperref}
\usepackage[normalem]{ulem}
\usepackage{xcolor}
\usepackage{color}
\usepackage{hyperref}

\hypersetup{colorlinks, citecolor=blue, linkcolor=red, urlcolor=blue}

\usepackage{graphicx,capt-of}

\newcounter{subfloat}

\def\be{\begin{equation}}
\def\ee{\end{equation}}
\newcommand{\dd}{\mathrm{d}\baselineskip0.5ex}
\def\Ms{{M_{\odot}}}
\def\xM{{x_{\rm max}}}
\def\xm{{x_{\rm min}}} 
\def\fnl{{f_{\rm NL}}}

\begin{document}

\title{NANOGrav signal as mergers of Stupendously Large Primordial Black Holes}

\author{Vicente Atal}
\email{vicente.atal@icc.ub.edu}
\author{Albert Sanglas}
\email{asanglas@icc.ub.edu}
\author{Nikolaos Triantafyllou}
\email{nitriant@icc.ub.edu}

\affiliation{Departament de F\'isica Qu\`antica i Astrof\'isica, i  Institut  de  Ci\`encies  del  Cosmos, Universitat de Barcelona, Mart\'i i Franqu\`es 1, 08028 Barcelona, Spain.}

\begin{abstract}
We give an explanation for the signal detected by NANOGrav as the stochastic gravitational wave background from binary mergers of primordial ``Stupendously Large Black Holes"(SLABs) of mass $M\sim(10^{11}-10^{12})\Ms$, and corresponding to  roughly $0.1\%$ of the dark matter. We show that the stringent bounds coming from $\mu$ distortions of the CMB can be surpassed if the perturbations resulting in these BHs arise from the non-Gaussian distribution of fluctuations expected in single field models of inflation generating a spike in the power spectrum. While the tail of the stochastic background coming from binaries with $M\lesssim 10^{11}\Ms$ could both fit NANOGrav and respect $\mu$ distortions limits, they become excluded from large scale structure constraints.
 \end{abstract}

\maketitle

\section{Introduction}

The NANOGrav collaboration has recently reported evidence for a signal consistent with a gravitational wave background of frequency  $\nu$ $\in (2.5\times10^{-9}$, $1.2\times 10^{-8})$ Hz and amplitude at 1-$\sigma$ confidence level $\Omega_{\rm GW} \in (3\times10^{-10}$,  $2\times10^{-9})$. If the signal is modelled as $\Omega_{\rm GW}\propto (\nu/\nu_{\star})^{\xi}$, the tilt $\xi\in(-1.5$, $0.5)$ at $\nu_{\star}=5.5$ nHZ and at 1-$\sigma$ confidence level \cite{Arzoumanian:2020vkk}.

The nature of the signal has still to be confirmed with further observations and analysis (e.g. whether it is really a gravitational wave (GW) and whether it is of stochastic origin or not), but since a potential GW detection might have tremendous consequences for our understanding of the Universe, it is important to investigate the potential implications of such discovery.
A suggestion of particular interest entails the signal being related to a population of primordial black holes (PBHs) \cite{Hawking(1971)}, which might be an important constituent of our Universe (for recent reviews, see e.g \cite{Sasaki:2018dmp,Green:2020jor}).
In this line, several possible interpretations of the NANOGrav signal  being the gravitational background induced by large scalar perturbations responsible for PBH formation has been proposed \cite{Vaskonen:2020lbd,DeLuca:2020agl,Kohri:2020qqd,Bian:2020bps,Sugiyama:2020roc,Domenech:2020ers,Bhattacharya:2020lhc,Inomata:2020xad}. In these works it has been shown that the signal could be accommodated by a population of sublunar, solar, or slightly supersolar PBHs.

A quite different possibility, although historically one of the first to be conceived for explaining a signal at such frequencies \cite{Rajagopal:1994zj}, is that it corresponds to the stochastic background resulting from the past mergers of large black holes, with masses $M>10^6 \Ms$. This possibility was studied in \cite{Arzoumanian:2020vkk,Middleton:2020asl} where the distributions of the binaries was inspired from astrophysical models.

Here we show that BH-inspirals accounting for NANOGrav could be of primordial origin. For this to happen, their mass should be $\sim10^{11}-10^{12}\Ms$, and so they enter in the class of so-called ``Stupendously Large Black Holes" (SLABs) \cite{Carr:2020erq}. Their abundance with respect to dark matter, $f\equiv \Omega_{\rm PBH}/\Omega_{\rm DM}$, should be at the $\mathcal{O}(0.1\%)$ level in order to explain the observed amplitude.

While primordial SLABs are heavily constrained by spectral distortions, we will show that a proper account of the non-Gaussianities (NG) arising in single-field inflation models leading to PBH production can easily relax these constraints. For these abundances, the lower limit on the mass, $M>10^{11}\Ms$, is obtained from dynamical constraints of large scale structure (LSS) (for a recent review on this topic, see \cite{Carr:2020erq}).

\section{Mergers of SuperMassive Black Holes and their Stochastic Gravitational Wave Signal}
The energy released by BH binaries contribute to a stochastic background of gravitational waves, whose energy density $\Omega_{\rm GW}$ is given by (we use $c=G=1$) \cite{Regimbau:2011rp}
\begin{align}\label{eq:OmegaGW}
\Omega_{\rm GW}&\equiv \frac{1}{\rho_c}\frac{d \rho_{\rm GW}}{d \log \nu} \nonumber \\
&=\frac{\nu}{\rho_c H_0}\int_{0}^{z_*}\frac{R_{\rm BH}(z)}{(1+z)E(z)}\frac{d E_{\rm GW}}{d \nu_s}(\nu_s)dz
\end{align}
where $\rho_{\rm GW}$ is the energy density at a given frequency $\nu$, $\rho_c$ is the critical density, $ \text{d}E_{\rm GW}/\text{d}\nu_s$ is the GW energy spectrum of the merger, $R_{\rm BH}$ is their merger rate at redshift $z$ and $\nu_s$ is the frequency in the source frame, related to the observed frequency as $\nu_s=(1+z)\nu$. The function $E(z)\equiv H(z)/H_0=[\Omega_r(1 +z)^4+ \Omega_m(1 +z)^3+ \Omega_\Lambda]^{1/2}$, and the GW energy spectrum of the merger is \cite{Ajith:2007kx}
\be\label{eq:dEdnu}
  \frac{d E_{\rm GW}}{d\nu_s}(\nu_s) = \frac{\pi^{\frac{2}{3}}M_c^{\frac{5}{3}}}{3}
  \begin{cases}
    \nu_s^{-1/3} &  \nu_s < \nu_1 \\
    \omega_1\nu_s^{2/3} & \nu_1 \leq \nu_s < \nu_2 \\
    \frac{\omega_2\sigma^4\nu_s^2}{\left(\sigma^2+4\left(\nu_s-\nu_2\right)^2\right)^2} &  \nu_2 \leq \nu_s < \nu_3 \\
    0 &  \nu_3 \leq \nu_s
    \end{cases}
\ee
\noindent where $\nu_i\equiv(\nu_1,\nu_2,\sigma,\nu_3)=(a_i \eta^2+b_i \eta + c_i)/(\pi M_t)$,  $M_t=m_1+m_2$ is the total mass of binary system, $M_c$ is the chirp  mass ($M_c^{5/3}=m_1m_2M_t^{-1/3}$),  and $\eta=m_1m_2 M_t^{-2}$ is the symmetric mass ratio. The parameters $a_i$, $b_i$ and $c_i$ can be found in \cite{Ajith:2007kx}, and ($\omega_1, \omega_2$) are chosen such that the spectrum is continuous.

The merger rate $R_{\rm BH}(z)$ depends on the formation channel of the binary system. If these are formed from primordial fluctuations,  then it will further depend on the their initial distribution, which might be Poissonian (if perturbations are Gaussian \cite{Ali-Haimoud:2018dau,Desjacques:2018wuu,Ballesteros:2018swv}), or clustered (if departures from Gaussianities are large \cite{Tada:2015noa,Young:2015kda,Belotsky:2018wph,Suyama:2019cst,Young:2019gfc,Atal:2020igj}).

\begin{figure}[tpb]
\centering
\includegraphics[width=\linewidth]{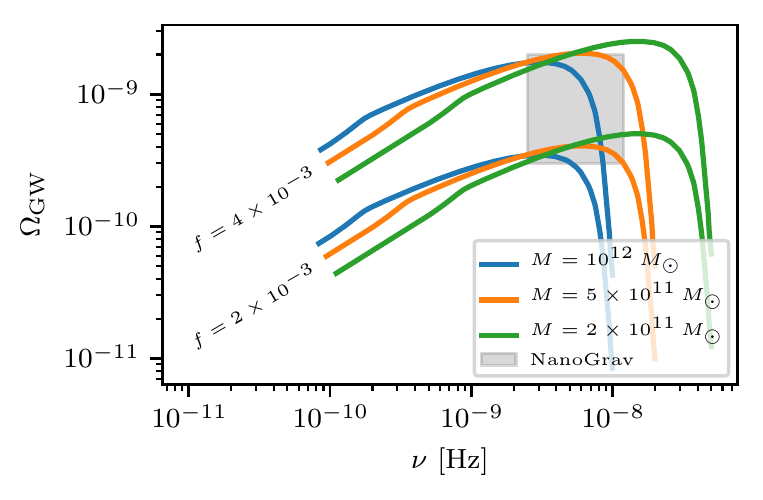}
\caption{The stochastic background of binary mergers. In grey we show the 1-$\sigma$ region consistent with NANOGrav signal.} \label{fig:OmegaGW}
\end{figure}

 While we will deal with non-Gaussian primordial fluctuations, the effect on their merger rate is going to be negligible since in the minimal model of inflation that we will consider here there is no large variance at large scales that can source a spatial modulation in the number density of PBHs. We consider thus a merger rate as coming from a Poissonian distribution of PBHs and a monochromatic mass function ($M\equiv m_1=m_2$), that is \cite{Nakamura:1997sm} \footnote{In Eq. (\ref{eq:mergerrate}) we have neglected an exponential factor that is irrelevant in this case but can be very important when large scale correlations are present \cite{Atal:2020igj}.}

\be\label{eq:mergerrate}
R_{\rm BH}(t)=8\pi^2\bar{n}^3\int_{\xm}^{\xM} x^2y^2 \Big|\frac{\dd y}{\dd t}(t,x)\Big|dx \ ,
\ee
\noindent where $\bar{n}$ is the comoving number density of BHs, $y = x\,(x/\xm)^{\frac{16}{21}}$, where $\xm$ and $\xM$ are the minimum and maximum distances of the binaries such that they merge at time $t$ \cite{Peters:1964zz},
\be\label{eq:xM}
\xM\simeq \left(\frac{M_t}{\rho_{eq}}\right)^{\frac{1}{3}} \quad \text{and}\quad \xm=\left(\frac{85\, \eta \, M_t^3 \,t}{3\, x^4_{\rm max}}\right)^{\frac{1}{16}}\xM \ ,
\ee
with $\rho_{eq}$ the background energy density at matter-radiation equality\footnote{Recently BHs described by the so-called Thakurta metric has been studied \cite{Boehm:2020jwd}. In that description, and due to a constant energy flow towards the BHs, their masses are time dependent, which alters their merger rate. Since in the cosmological setup that we are interested there is no fluid sustaining this mass grow, these BHs are not the ones that we are interested in (see also \cite{DeLuca:2020jug}).}.
In Fig. \ref{fig:OmegaGW} we show how the NANOGrav signal can be fitted with the stochastic GW signal given by Eq. (\ref{eq:OmegaGW}). It can be attributed to the peak of the signal if PBHs are of masses $M\sim(5\times10^{11}-10^{12})\Ms$ and make a fraction $f\sim(2-4)\times10^{-3}$ of the total DM. The infrared tail of the stochastic background has a spectral index $\xi=2/3$, and is thus  within the 2-$\sigma$ interval determined by observations. Thus BHs of $M<5\times 10^{11}\Ms$ and $f > 2\times10^{-3}$ could also provide a good fit. However, as we will see in the next section, these become in conflict with LSS constraints. In Fig. \ref{fig:OmegaGW} we show the smallest mass that provides a fit to NANOGrav that is not in tension with LSS constraints. The combination of both constraints imply $M\sim(2\times10^{11}-10^{12})\Ms$ and $f\sim(2-4)\times10^{-3}$.  

While such large BHs have yet to be observed in nature (the largest reported BH has a mass $7\times 10^{10} \Ms$ \cite{Shemmer:2004ph}), the presence of more massive BHs is not ruled out\footnote{For for astrophysical considerations on their presence, see \cite{Natarajan:2008ks}.}. In the following we discuss limits on the fraction of PBHs of this range of masses, most notably the spectral distortion and large scale structure bounds.

\section{Spectral distortions, Non-Gaussianities and LSS}

Spectral distortions provide the most stringent bound for the presence of PBHs in the mass range $M=(10^4-10^{12})M_{\odot}$. The large fluctuations necessary for the production of PBHs dissipate through Silk damping and might leave a large imprint in the CMB as departures from a black-body spectrum\footnote{A natural exception are models in which the PBHs are not associated to an enhancement of the power spectrum (like \cite{Garriga:2015fdk,Deng:2016vzb}). These models are thus essentially unconstrained by spectral distortions measurements.}. In particular, these primordial inhomogeneities are constrained by a non-detection of $\mu$ distortions, which from COBE/FIRAS is bounded to be smaller than $9\times 10^{-5}$ \cite{Fixsen:1996nj}.

The amplitude of the $\mu$ distortions depends on both the scale and the variance of the power spectrum of the scalar perturbations \cite{Chluba:2012we}. Both of these are related to the abundance of PBH, and so constraints on $\mu$ distortions can be used to constraint the presence of PBHs. From here it has been deduced  that PBHs in the mass range $(6\times 10^4-5\times 10^{13})\Ms$ are excluded \cite{Kohri:2014lza} (for earlier application of this idea, see \cite{Carr:1993aq,Carr:1994ar}).

These constraints can be relaxed if NG are considered \cite{Nakama:2017xvq,Unal:2020mts}. The abundance of PBH is most sensitive to the ratio $\nu\equiv\zeta_c/\sigma$, and NG might change both $\sigma$, the variance of the perturbations, and $\zeta_c$, the critical threshold for collapse. In general the role that NG plays in determining the threshold for collapse has been neglected, and only the effect coming from changes in the PDF has been considered \cite{Nakama:2017xvq,Unal:2020mts,Inomata:2020xad}.
However, the threshold for collapse depends on the shape of the overdensity \cite{Shibata:1999zs,Harada:2015yda,Germani:2018jgr,Musco:2018rwt,Yoo:2018kvb,Escriva:2019nsa,Escriva:2019phb}, and non-Gaussianities modify its shape \cite{Atal:2019cdz,Yoo:2019pma,Kehagias:2019eil,Atal:2019erb}. 

For example, if we consider local models of NG, for which the curvature perturbations $\zeta$ are related to a Gaussian variable $\zeta_G$ with a local function $\zeta=F\left(\zeta_{G}\right)$,  the profile of an overdensity is simply $F(\zeta_G(r))$, where $\zeta_G(r)$ is its shape as given by a Gaussian random field \cite{Bardeen:1985tr}. Then the treshold for collapse can be simply determined numerically (with public codes \cite{Escriva:2019nsa}) and/or analytically \cite{Escriva:2019phb}.

Note that since local models of NG can be written in terms of an underlying Gaussian field, it is not necessary to find how the PDF changes by the local transformation for computing the abundances. It is actually sufficient to count the regions for which the image of the underlying Gaussian field is above the threshold of collapse. That is, if $\zeta^{c}_{\textrm NG}$ is the critical value for collapse of the NG profile, we can define its counterpart in the underlying Gaussian field, $\mu_c$, given by $\mu_c\equiv F^{-1}(\zeta^{c}_{\textrm NG})$.
Then, if the PBH abundance  for a Gaussian field is $\beta_G \sim e^{-\left(\zeta^c_{\textrm G}\right)^2/\sigma_G^2}$ with $\zeta_{G}^{c}$ the treshold for the Gaussian field, then the abundance of PBHs in the local NG theory is simply given by $\beta_{NG}\sim e^{-\mu_c^2/\sigma_G^2}$. 

It is thus fortunate that in the case of single-field inflationary models producing a spike in the power spectrum (necessary for a controled production of PBHs), the NGs are of the local type. Its precise shape and amplitude were established in \cite{Atal:2019cdz}, and are given by
\begin{equation}\label{eq:NG_log}
\zeta=-\mu_*\log\left(1-\frac{\zeta_g}{\mu_*}\right)
\end{equation}
\noindent where $\mu_*$ is related to the curvature of the local maxima that the inflaton field traverses, as \cite{Atal:2018neu,Atal:2019cdz}
\be\label{eq:mu_c_V}
\mu_*=-3+\sqrt{9-\eta_V|_{\rm max}}
\ee
\noindent where $\eta_V\equiv V''/V$, with $V(\phi)$ the inflaton potential, and where $'$ denotes derivatives with respect to inflaton field $\phi$. The parameter $\eta_V$ in Eq. (\ref{eq:mu_c_V}) is to be evaluated at the local maxima of the potential. As the field overcomes the local maxima it enters a stage of constant-roll, perturbations are largely amplified and the abundance of PBHs increases.
The local transformation Eq. (\ref{eq:NG_log}) is a non-perturative completion of the well known perturbative version $F(\zeta_G)=\zeta_G+(3/5)f_{\rm NL}\zeta^2_{\rm G}$ \cite{Atal:2019erb}. Thus, we can relate the parameter $\mu_*$ to $\fnl$, as
\be
\mu_*=\frac{5}{6\fnl} \ .
\ee
The local transformation Eq. (\ref{eq:NG_log}) induces a change in the  shape of the overdensity field with respect to a Gaussian field. In \cite{Atal:2019erb} it was determined how the threshold for collapse changes as the parameter $\mu_*$ (or equivalently $\fnl$) varies. For a peaked power spectrum growing as $k^4$, as expected in single-field inflationary models \cite{Byrnes:2018txb}, we find\footnote{Actually there are no large differences at the level of the threshold if we had considered a power spectrum given by a $\delta$ function. For details, see \cite{Atal:2019erb}.}
\be
\mu_{\rm c}=
\begin{cases}
0.663-0.183\fnl+0.0169 f^2_{\rm NL} & \fnl<5\\
5/(6\fnl) &\fnl>5 
 \end{cases}
\ee
Thus, as the amplitude of the NG is increased (which in terms of the potential means that the hill that the inflaton must traverse is more and more spiky), the threshold for collapse diminishes. Note that in the non-perturbative template described here the change in the threshold is more important than in its perturbative version $F(\zeta_G)=\zeta_G+(3/5)f_{\rm NL}\zeta^2_{\rm G}$ \cite{Atal:2019erb}. Also, since the threshold is found here in terms of $\zeta$, we avoid complications related to additional NGs from the non-linear relation between curvature and density perturbation. 
 
For determining the spectral distortions it is necessary to calculate the variance of the non-Gaussian field\footnote{This turns out to be very close to the Gaussian variance, since the logarithm in Eq. (\ref{eq:NG_log}) only affects profiles close to $\mu_*$, which are very rare with respect to the typical profiles of $\zeta$.}. This is given by
\begin{equation}
\sigma^2=\frac{1}{\sqrt{2\pi\sigma_0^2}}\int_{-\infty}^{\mu_*}\log\left(1-\frac{\zeta_g}{\mu_*}\right)^2e^{-\frac{\zeta_g^2}{2\sigma^2_0}}\,d\zeta_g - \left<\zeta\right>^2  \ .
\end{equation}
The amplitude of the $\mu$ distortion is related to the amplitude and scale of a perturbation as \cite{Chluba:2012we}
\be
\mu\simeq2.2\sigma^2\left[\exp\left(-\frac{k_\star}{5400}\right)-\exp\left(-\left(\frac{k_\star}{31.6}\right)^2\right)\right] \ .
\ee
Here the wavenumber $k_{\star}$ is in $M_{\rm pc}^{-1}$. Assuming a monochromatic mass function -a good approximation for peaked power spectra- $k_{\star}$ is related to the mass of the PBH as \cite{Nakama:2016gzw}
\be
k_{\star}\simeq7.5\times 10^5\gamma^{\frac{1}{2}}\left(\frac{g}{10.75}\right)^{-\frac{1}{12}}\left(\frac{M}{30\Ms}\right)^{-\frac{1}{2}} \ .
\ee
Here $\gamma$ is the ratio between the mass of the BH  and the mass enclosed within the horizon $H=ak^{-1}$, and $g$ is the number of relativistic species (we fix $g=10.75$). For simplicity we take $\gamma=1$, knowing that we expect some departures from unity from the fact that the BH collapse is a critical phenomena \cite{Niemeyer:1999ak}, and because the size of the overdensity undergoing the collapse is typically some factors larger than $a k^{-1}$ \cite{Germani:2018jgr,Yoo:2018kvb}. 

\begin{figure}[tpb]
\centering
\includegraphics[width=\linewidth]{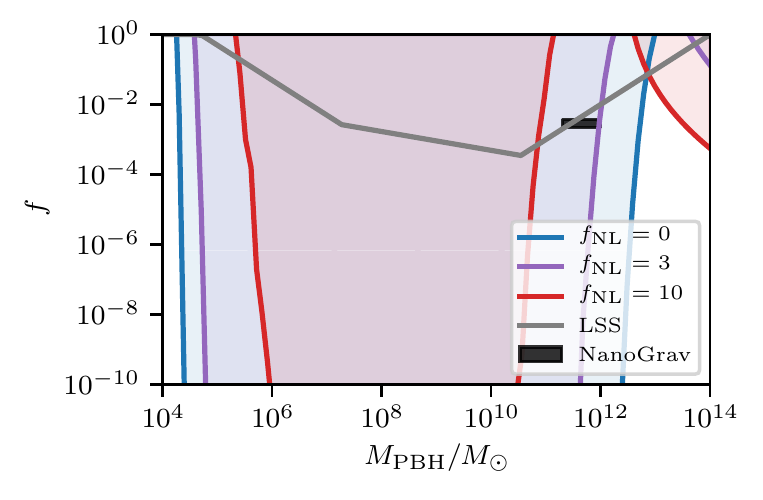}
\caption{Limits on the abundance of PBHs coming from the non observations of $\mu$ distortions in the CMB and from large scale structure considerations. We can see that for $f_{\rm NL}\gtrsim3$, PBHs with the mass and abundance appropiate to fit NANOGrav are not constrained by $\mu$ distortions. However, for these abundances, the formation of clusters imposes $M>2\times 10^{11}\Ms$. The region in black corresponds to the fits shown in Fig. \ref{fig:OmegaGW}}. \label{fig:Distortion_Limit}
\end{figure}

The fraction of PBH to DM, $f$, is related to their primordial abundance as \cite{Nakama:2016gzw}
\be
\beta\simeq 10^{-8}\gamma^{-\frac{1}{2}}\left(\frac{g}{10.75}\right)^{\frac{1}{4}}\left(\frac{\Omega_{\rm DM}}{0.27}\right)^{-1}\left(\frac{M}{30\Ms}\right)^{\frac{1}{2}}f \ .
\ee
The abundance can be calculated using the standard peak theory \cite{Bardeen:1985tr}, since we mention previously that we can always refer to the underlying Gaussian field\footnote{It has been shown that computing the abundances using the compaction function gives a similar result for peaked power spectra \cite{Germani:2019zez}.}. For the case of a monochromatic power spectrum, it is simply given by \cite{Germani:2018jgr,Yoo:2018kvb}
\be
\beta\simeq\frac{\mu_c^3}{\sigma_0^4}\exp\left(-\frac{\mu_c^2}{2\sigma_0^2}\right) \ .
\ee
Here $\mu_{c}$ is the threshold of the underlying Gaussian field, and $\sigma_0$ its variance. In fig. \ref{fig:Distortion_Limit} we show the constrains from the FIRAS limit on $\mu$, and how they vary with increasing non-Gaussianity parameter $\fnl$. For $\fnl>3$ these constraints can be avoided. Interestingly, $\fnl>3$ sets the limit above which most PBHs are baby Universes, resulting from trapped regions in the false vacuum of the potential \cite{Atal:2019erb}.   
Let us note that by considering a power spectrum growing as $k^4$, we should also worry about the constraints at scales larger than the peak of the power spectrum, since those are also constrained by the bounds on $\mu$ distortion. In this respect, we should note that by virtue of a duality between the background before and after the peak in the power spectrum, the amplitude of the non-Gaussianity remains constant during the transition \cite{Kinney:2005vj,Atal:2018neu}.
Thus, the most critical test concerning $\mu$ distortions is at the scale of the peak, $k_\star$, irrespective of the steepness of the power spectrum. Note also that in similar inflationary scenarios than the one we describe here there might be an additional source of $\mu$ distortions, coming from shock waves of an expanding bubble of false vacuum in the thermalized fluid \cite{Deng:2018cxb}. In our case, when the BH corresponds to regions in the false vacuum ($\fnl>3$), the energy of the false vacuum is larger than the background energy density and so the bubble does not expand from the point of view of an outside observer and thus there are no such shock waves.

Limits from $\mu$ distortions are not the only ones constraining PBHs of these masses. The effects of gas accretion in the thermal history of the Universe might also be important. For this range of masses, their effect was studied in \cite{Carr:1981mnras}. In the simplified model considered there, for an efficiency $\epsilon<0.03$, $f\sim 10^{-3}$ is allowed. We note that these constraints should be updated considering more recent studies, that however mostly focused in $M<10^4\Ms$ (see e.g. \cite{Ricotti:2007au,Ali-Haimoud:2016mbv}).

We can also consider dynamical constraints. Massive PBHs could destroy galaxies in clusters \cite{Carr:1997cn} and/or trigger an undesirably early  formation of cosmic structure \cite{Carr:2018rid}. In Fig. \ref{fig:Distortion_Limit} we show the limits coming from the formation of dwarf galaxies up to clusters of galaxies. For example, in the mass range $M\sim(3\times10^{10}-10^{14})\Ms$, the strongest limit comes from the formation of clusters of galaxies, and reads \cite{Carr:2018rid}
\be
f<\frac{M}{10^{14}\Ms} \quad\text{for}\quad 3\times10^{10}\Ms<M<10^{14}\Ms \ .
\ee
These constraints are fully consistent with fitting NANOGrav with the peak of the stochastic background, but put limits on the fits coming from its tail, fixing $M>2\times10^{11} \Ms$. All in all, these constraints imply that the NANOGrav signal can be explained if  $M\sim(2\times10^{11}-10^{12})\Ms$, $f\sim(2-4)\times 10^{-3}$ and $\fnl>3$. Note that for these parameters the signal is no longer well described by a power law in the relevant range of frequencies, and thus motivates an extension of the templates used in \cite{Arzoumanian:2020vkk,Middleton:2020asl} to asses the significance of the fits.

\section{Conclusions}

We have provided an explanation of the NANOGrav signal as coming from the binary mergers of Stupendously Large Black Holes, of masses in the range $M\sim (2\times10^{11}-10^{12}) \Ms$. For a mild non-Gaussianity resulting from the dynamics of single field inflation, the spectral distortions constraints can be avoided. For the abundance needed for explaining NANOGrav, $f\sim 10^{-3}$, the dynamical constraints fix their mass to be $M>2\times 10^{11}\Ms$.

These results provide yet another motivation for studying the presence of SLABs, and illustrate the importance of non-Gaussianities and their connection to the formation mechanism of PBHs for determining bounds on their presence.

Finally, we should indicate that in this work we have neglected the mass evolution of the PBHs. Since most of the signal from the stochastic background comes from late mergers, the mass evolution could be important (see e.g. \cite{Garcia-Bellido:2017aan}). Then it would be necessary to invoke PBHs of smaller masses for having a late time signal at the frequencies of NANOGrav. While large  values of $\fnl$ can easily be obtained to surpass the $\mu$ distortion limits \cite{Atal:2019cdz}, bounds on LSS are more difficult to overcome. Considering that at the moment bounds from LSS and gas accretion are only order of magnitude constraints for such masses, a deeper examination of them should be pursued for a more robust model selection.

\begin{acknowledgments}
We thank Jaume Garriga for comments on the manuscript. This work has been partially supported by FPA2016-76005-C2-2-P, MDM-2014-0369 of ICCUB (Unidad de Excelencia Maria de Maeztu), and AGAUR2017SGR-754 (V.A), by an APIF grant from Universitat de Barcelona (A.S), and by an INPhINIT grant from laCaixa Foundation (ID 100010434) code LCF/BQ/IN17/11620034 (N.T).  This project has also received funding from the European Unions Horizon 2020 research and innovation programme under the Marie Sklodowska-Curie grant agreement No.  713673. 
\end{acknowledgments}

\bibliographystyle{plain}

\end{document}